\documentclass[onecolumn]{elsarticle}
\usepackage{caption}
\pdfoutput=1
\usepackage{graphicx}
\usepackage {subcaption}
\usepackage{float}
\usepackage{subfloat}
\usepackage{natbib}
\usepackage{geometry}
\usepackage{txfonts}
\usepackage{endfloat}
\usepackage{booktabs,caption,fixltx2e}
\usepackage{caption}

\begin{document}
\title{Dark Matter -- Modified Dyanamics:  Reaction vs. Prediction}
\author[1]{Robert H. Sanders}
\address{Kapteyn Astronomical Institute, Groningen, The Netherlands}

\begin{abstract}
The dark energy-cold dark matter paradigm ($\Lambda$CDM) has gained widespread
acceptance because it explains
the pattern of anisotropies observed in the cosmic microwave background
radiation, the observed distribution of large scale inhomogeneities
in detectable matter, and the perceived overall expansion history of the Universe.
It is further {\it assumed} that
the cosmic dark matter component clusters on the scale of bound astronomical systems and thereby
accounts for the observed difference between the directly detectable
(baryonic) mass and the total Newtonian dynamical mass.  In this respect the
paradigm fails;  it is falsified by the existence of a simple algorithm,
modified Newtonian dynamics (MOND), which explains, not only general scaling
relations for astronomical systems, but quite precisely predicts
the effective gravitational acceleration in such objects from the observed
distribution of detectable baryonic matter -- all of this with one additional
universal parameter having units of acceleration.  On this sub-Hubble scale,
the dark matter
hypothesis is essentially reactive, while MOND is successfully predictive.
\end{abstract}
\maketitle

\section{Introduction}
The world, we are told with great certainty, is composed of 68.3\%  dark energy,
26.8\% dark matter and only 4.9\%
of the familiar directly observable baryons.  This description of
the Universe is recited at the beginning of almost all lectures
on cosmology -- like a catechism -- which is appropriate because
it is almost an article of faith.  And yet, we have no clear idea
what these two dominant substances actually are but only the general properties
they must have in order to meet the observational requirements
(see Sanders 2016 and references therein for a general
discussion of the modern cosmological paradigm, also Merritt 2017).

Dark energy is mysterious.  The direct empirical evidence for
its existence is the observed acceleration in the expansion
of the Universe seen in the radial velocity of high redshift
(distant) galaxies.  Because gravity is normally attractive this
can be taken to be a cosmological
term in Einstein's field equations which appears
as a repulsive force. This works but leaves one
feeling somehow unsatisfied as an explanation.  The cosmological term ($\Lambda$)
can be interpreted
as an energy density of the vacuum --  an additional (negative) source term on
the right hand side of the field equations --  and as such can be given
an equation of state, a fluid with pressure equal to negative energy density
($p=-\rho$) that provides repulsion and guarantees its constancy with the expansion of the
Universe; i.e., this fluid does not dilute with volume as the Universe expands.
But then the value of the density seems unnaturally small,
particularly with respect to the matter-energy density of the Universe
near the Planck epoch
($10^{-121}$ give or take a order of magnitude).
It could be the energy density of an evolving field
allowing a more general equation of state ($p = w\rho$ where $w<-1/3$) and
therefore having some dilution with expansion.
But then what is this new cosmic field and how does it relate to physics
in general?  It seems disturbing that the major constituent of the Universe,
the only evidence for which is astronomical, is
not understood.

The second major component, dark matter, seems more comprehensible.
We can all imagine particles like ping-pong balls bouncing around the Universe, diluting
with the expanding volume, and
contributing through their mass density the necessary attraction (via gravity)
to counterbalance
(almost at present) the expansion driven by dark energy. Dark matter is required on a
cosmological scale in
order in order to understand the observed expansion history of the Universe --
in particular the transition from deceleration, where matter dominates,
to acceleration at a redshift of about one, where dark energy begins to dominate.
And then, it is required in order to form the observed structure --
galaxies,
clusters of galaxies, enormous filaments and voids of galaxies -- in the
necessary finite
time via gravitational collapse in an expanding Universe.  And finally
there is direct evidence in the pattern of anisotropies in the cosmic
microwave background which are consistent with photon-baryon oscillations
in the pre-recombination universe
in the presence of more rigid dark matter concentrations.

In fact, the original motivation for dark matter is the discrepancy between
the directly observed mass in baryons -- stars and gas -- and the Newtonian
dynamical mass of self-gravitating systems -- galaxies
and clusters of galaxies -- estimated by size and internal velocity.
That is to say, there is a local motivation
for dark matter that is present on scales much smaller than the Universe as a whole,
the Hubble scale $c/H_0$.
and that means dark matter must cluster at least on the scales of galaxies.  This
requires that considered as a fluid it must be cold -- cold dark matter or
CDM -- which is to say when it decouples from the rest of the Universe at
early epochs its velocity dispersion is less than the speed of light.
So this cosmological substance
impinges directly upon dynamics of these local systems.
Because it clusters locally, this leads to the expectation that the particles
may be detectable locally,
even in terrestrial experiments. Thus far, in spite of heroic efforts,
dark matter particles have not been seen in any non-astronomical experiment, meaning
that the particles must interact very weakly with ordinary baryonic matter and
with themselves.  These mysterious particles cannot be charged,
and they must be stable on cosmic timescales.  There is no standard
model particle which meets these requirements with the possible exception
of neutrinos (and these should be non-standard neutrinos).
The primary interaction must be gravitational
just as the primary evidence at present is astronomical.

And this brings us to the essential motivation for an alternative paradigm:
{\it When a theory (in this case general relativity), extended into a regime
where it has never before been tested (low acceleration systems), requires
a medium, an aether (dark matter-dark energy), that cannot be detected by any means
independent of the phenomena it is introduced to explain
 -- then it is not unreasonable to question that theory}.
There is such an alternative, modified Newtonian dynamics,
which here I will view as a simple algorithm that allows the
the distribution of force in an astronomical object to be
calculated from the observed distribution of baryons with one
new universal parameter having units of acceleration.
These predictions for spiral galaxies work very well as evidenced
by the matching of calculated with observed rotation curves
in these objects, in a number of case with no free parameters
apart from the universal acceleration.  This fact would appear
totally at odds with dark matter as it is perceived to be:  a
dissipationless fluid that interacts with normal matter only
via gravity.

\section{Modified Newtonian Dynamics:General Predictions (Milgrom 1983)}

Milgrom noticed more than 30 years ago that in bound astronomical systems, the
discrepancy between directly observable mass and the classical
dynamical mass does not appear preferentially in large systems but in low
acceleration systems.  Thus he introduced a modification of
Newtonian dynamics not connected to a length scale but to 
an acceleration scale.  And this algorithm has turned out to provide, in
its most primitive form.
a very efficient summary of galaxy phenomenology.  Moreover
it is predictive.  New phenomena not seen earlier were accurately
predicted before being observed.  And it is simple -- as simple
as $F=ma$:
$$g\mu(a/a_0)= g_n  \eqno(1))$$
where $g$ is the ``true'' acceleration, $g_n$ is the traditional
Newtonian acceleration calculated for an observed mass distribution
via the traditional Poisson equation, $a_0$ is the
single new universal parameter, and $\mu(x)$ is a function
of the acceleration in terms of $a_0$ which interpolates
between the high acceleration Newtonian regime and the 
phenomena at heretofore unexplored low acceleration.
This function is not specified but must have definite asymptotic
behaviour: $\mu(x) = 1$ when $x>>1$ and $\mu(x)= x$ when $x<<1$.

\begin{figure}
\includegraphics[height=10cm]{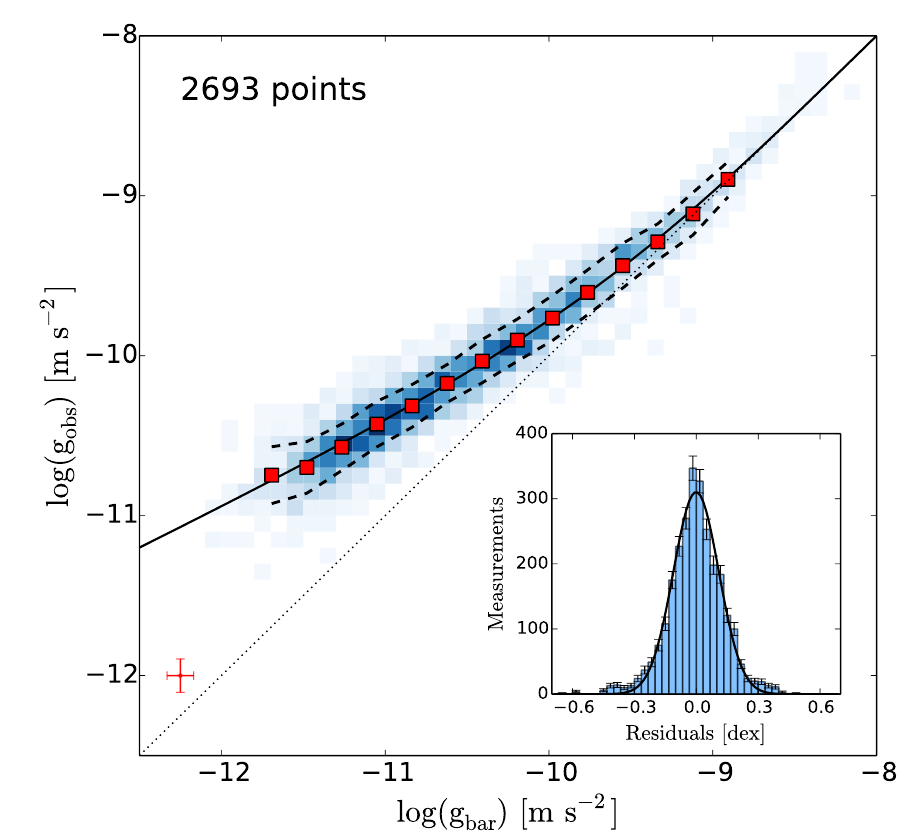}
\captionsetup{margin=10pt,font=small,labelfont=bf}
\caption{The radial acceleration relation (RAR).  This is a plot
of the measured (true) acceleration ($g$) against the Newtonian
acceleration of the baryonic component over a range of radii
in aprroximately 100 spiral galaxies
It is essentially $g$ vs $g_n$ in eq.\ 1.  The shape of the curve
defines the interpolating function, $\mu(g/a_0)$ (see McGaugh, Lelli
\& Schombert 2014)
)
}

\end{figure}

The MOND equation forms the basis for the recently emphasised
radial acceleration relation (McGaugh et al. 2016) as shown in Fig.\ 1.
This is a plot of the measured centripetal acceleration in spiral galaxies
against the Newtonian acceleration of the baryonic
mass distribution, stars and gas, in a sample of about 100 galaxies over a
range of radii.
It is a perfect demonstration of the validity of eq.\ 1 in the real
world and the role of a critical acceleration in differentiating
between the Newtonian and non-Newtonian behaviour.  The asymptotic
regimes are clearly evident and the shape of this curve defines
the interpolating function $\mu(a/a_0)$.

At high accelerations the true acceleration is equal to
the Newtonian acceleration, but at low accelerations
$$g = \sqrt{a_0 g_n} \eqno(2))$$
In the very low acceleration limit ("deep MOND") the effective
force about an isolated point mass $M$ becomes
$$g = \sqrt{GMa_0}/r \eqno(3)$$
which is to say, the asymptotic
circular velocity ($V^2/r$) is $$V=({GMa_0})^{1/4} . \eqno(4)$$
The rotation velocity about an isolated mass is asymptotically constant
at a value proportional to the one-fourth power of the mass.  In other
words, asymptotically flat rotation curves and a baryonic mass -- rotation velocity
(Tully-Fisher) relation are subsumed by this idea.

Now one might argue that these two aspects of galaxy phenomenology
are not predictions because they are part of the axiomatic basis
or MOND, but in several major aspects they are true predictions.
MOND as a modification of
physical law means that every isolated galaxy should exhibit
an asymptotically flat rotation curve and all such objects should fall
on the same mass-rotation velocity relationship -- without exception.
The velocity entering this relation and minimises its scatter
is the asymptotic constant velocity.
Moreover, the parameter $a_0$ that enters the rotation curve is the
same constant that normalises the Tully-Fisher relationship.
These are predictions and they are verified by more than 100
galaxies such as that in Fig.\ 2 and by the observed baryonic
Tully-Fisher relation shown by Fig.\ 3 (McGaugh 2011).

\begin{figure}
\includegraphics[height=10cm]{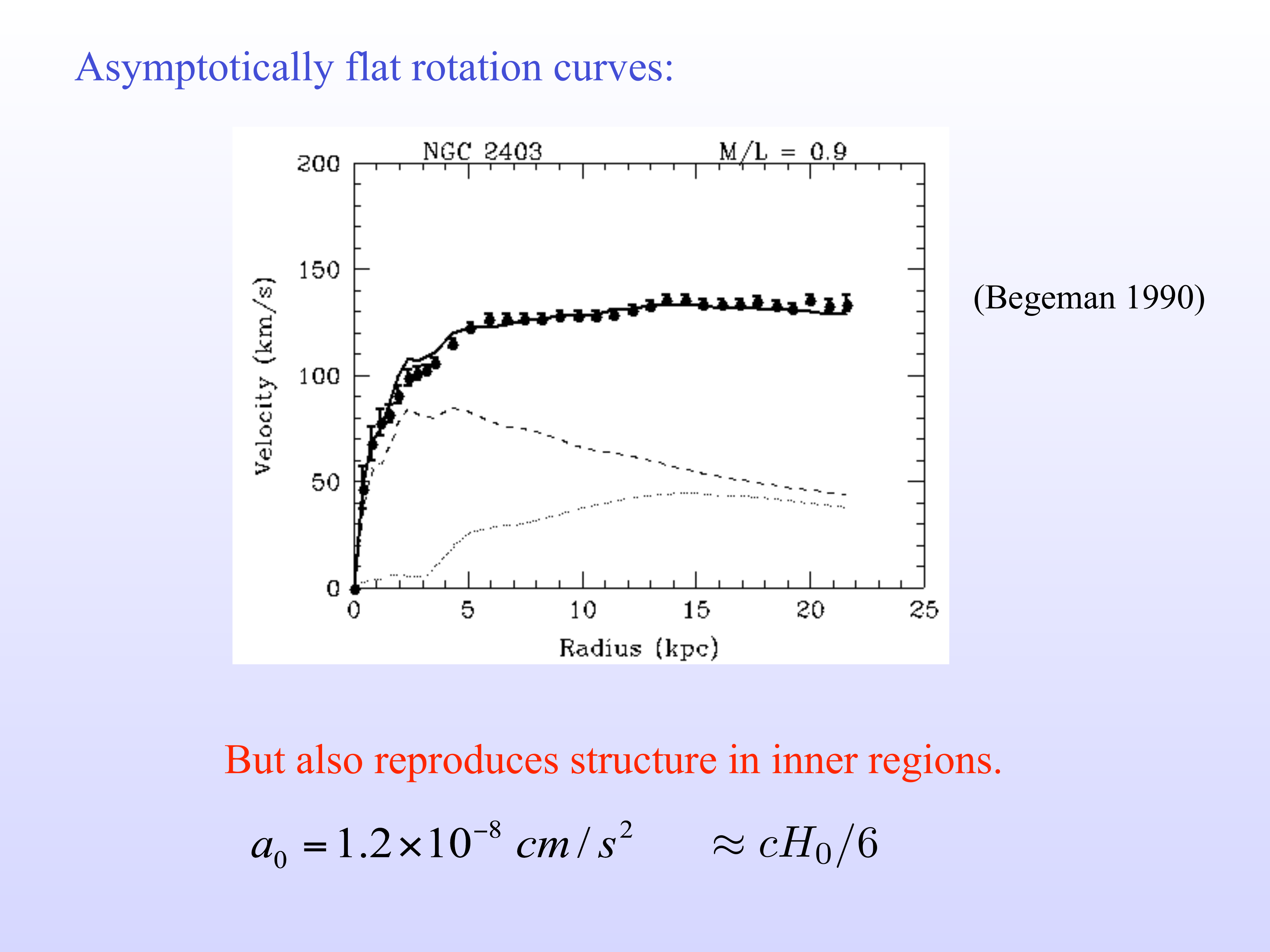}
\captionsetup{margin=10pt,font=small,labelfont=bf}
\caption{Rotation curve (rotation velocity vs. radius) of NGC 2403 as measured in the 21 cm
line of neutral hydrogen.  The points are the observations, the dashed curve is th
Newtonian rotation curve of the stellar component, the faint dashed curve is the
Newtonian rotation curve of the gaseous compoent, and the solid curve is the
MOND rotation curve (Begeman et al. 1991).
}

\end{figure}

\begin{figure}
\includegraphics[height=10cm]{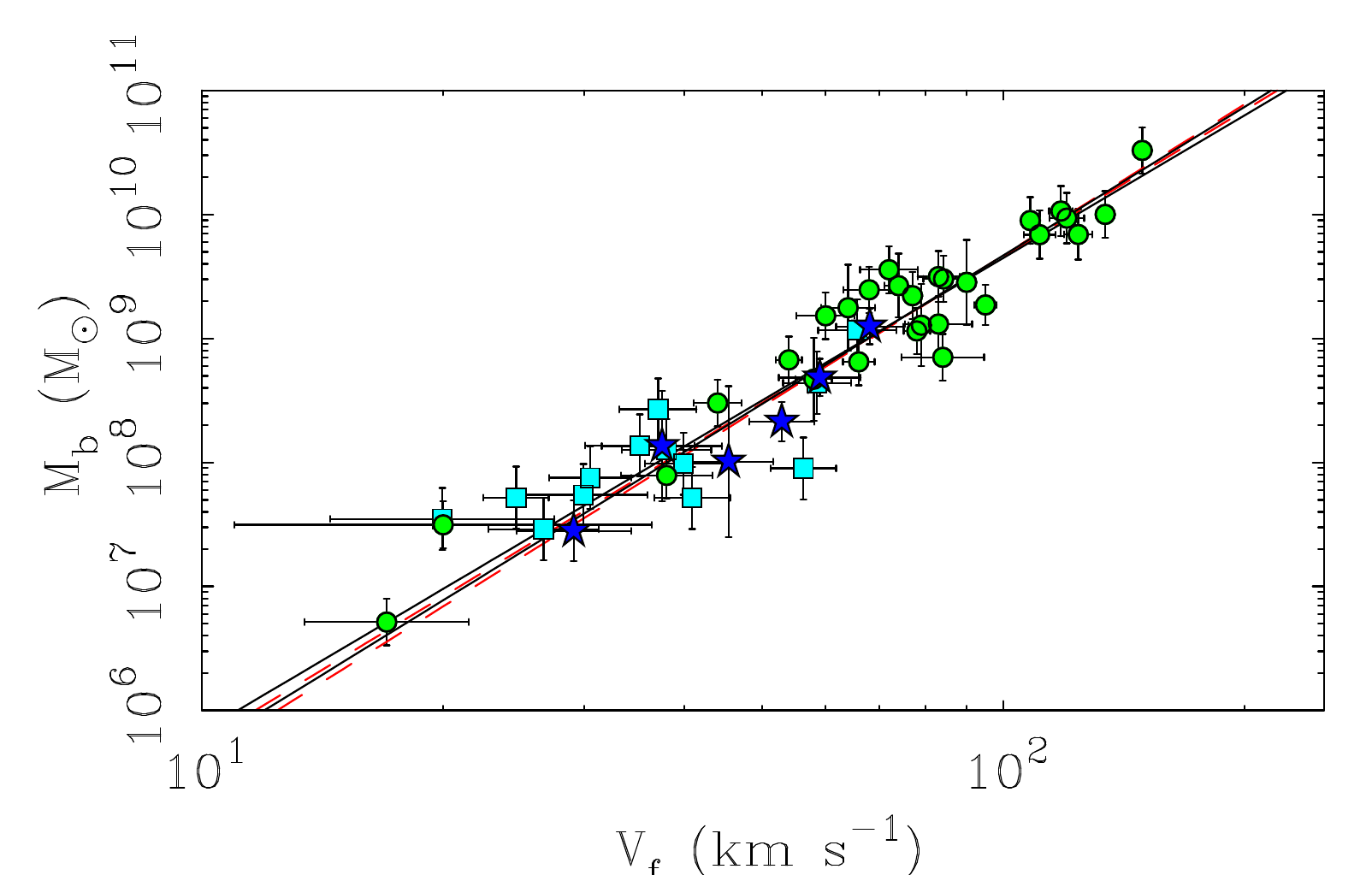}
\captionsetup{margin=10pt,font=small,labelfont=bf}
\caption{
Baryonic Tully-Fisher relationship.  Asymptotic
rotation velocity plotted against baryonic (detectable) mass
(McGaugh 2010)
}
\end{figure}

The acceleration parameter, empirically determined from rotation curves
and the Tully-Fisher relation, has the cosmologically significant value
of $a_0 \approx cH_0/6$, which suggests a connection between cosmology
and dynamics of local systems.  There has been much
work over three decades attempting to formulate a deeper theory of MOND
beginning with the Bekenstein \& Milgrom (1984) non-relativistic Lagrangian-based
modification of gravity,
but here I will consider
MOND as the original simple algorithm in terms of its predictive power
(see Famaey \& McGaugh 2012, Milgrom 2014).

\begin{figure}
\includegraphics[height=10cm]{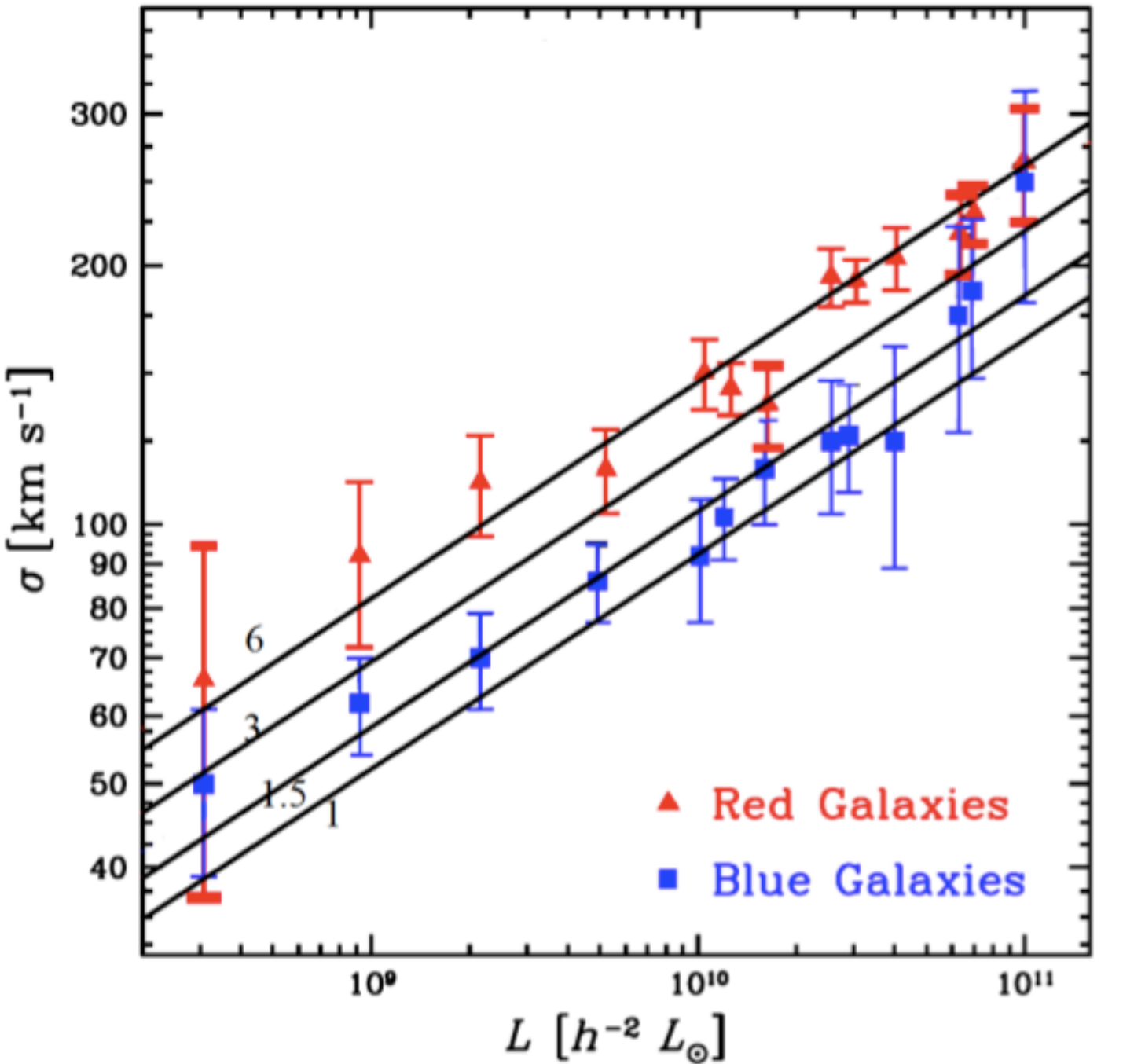}
\captionsetup{margin=10pt,font=small,labelfont=bf}
\caption{Extended
Baryonic Tully-Fisher relationship (Milgrom 2013).  The solid lines show
the MOND predictions for the $\sigma$-luminosity relation
for the indicated mass-to-light ratios.  The points are
the results of weak gravitational lensing analysis of isolated
foreground galaxies (blue and red).
}
\end{figure}

An extension of the Tully-Fisher relation has been recently pointed out by Milgrom
(2013).
It is now possible to apply weak gravitational lensing (systematic distortions of
background galaxies in the presence of an isolated foreground galaxy)
to statistically trace the
effective force distribution well beyond the visible image of
the foreground galaxy in relatively gas-free systems (Brimioulle et al. 2013).
A Newtonian isothermal
sphere (with $\rho = \sigma_r^2/(2\pi G r^2$) fits the resulting mass distributions (presumably the dark halo)
reasonably well, and one
finds that the implied velocity dispersion (proportional to the asymptotically constant circular
rotation velocity) bears the same relationship to the baryonic mass of the central
object as that implied by MOND, i.e., eq.\ 4.  This is true for radial
distances out to  several hundred kpc or accelerations on the order $0.01\,a_0$.
Milgrom's result is shown in Fig.\ 3 which shows the fitted velocity
dispersion of the isothermal sphere as a function of the luminosity of
the central galaxy -- a proxy for the baryonic mass but one dependent
upon the mass-to-light ratio and this depends upon colour with redder
galaxies having a larger M/L (this result has been confirmed to higher
precision by Brouwer et al. 2017).
In terms of dark matter this would mean that the dynamics of the
extensive dark halo well beyond the visible disk is correlated with the mass of
a trace of
visible matter in the very central regions which would seem quite
extraordinary in the context of dark matter.  
With MOND this is the expected result.

Milgrom in his original papers made
an essentially new prediction that was not anticipated
before MOND concerning the existence of a critical
surface density or surface brightness.
This arises because the MOND acceleration constant
$a_0$ can also be written as a surface density,
$$\Sigma_c = a_0/(\pi G). \eqno(5)$$
With the standard value of $a_0$ the numerical value is 0.19 $g/cm^2$ or 270
$M_\odot/pc^2$. With a typical mass-to-light ratio of
one to two in solar units, this would (in the peculiar astronomical
units) correspond to a critical
surface brightness on the order of 22 $mag/arcsec^2$
Because systems with a surface brightness greater than
this limit are in the Newtonian regime (high surface brightness 
systems such as globular star clusters or luminous ellipticals)
this would imply that such objects should exhibit a small
mass discrepancy within the visible object (little need for dark
matter).  In the other limit extreme low surface brightness systems
(dwarf spheroidal, low surface brightness spiral galaxies), should
exhibit a large discrepancy within the visible object.

This prediction has been consistently verified by subsequent
observations.  There were very few kinematic observations
of dwarf spheroidal galaxies or LSB spirals
in 1983 when Milgrom made his prediction,
but now there are many -- and all have large mass discrepancies (de Blok \& McGaugh 1998).
There also have now been detailed spectroscopic observations of
luminous ellipticals using bright tracers such as planetary nebulae.
It came as a surprise to many
when these objects were found to require little dark matter within the
visible object (Romanowsky et al. 2003, Milgrom \& Sanders 2003).

Recently there have been claims that several large LSB galaxies exhibit
a small discrepancy between visible and dark matter within the
optical image (van Dokum et al. 2018, Mancera Pi\~na et al. 2019).
If there are such isolated, virialized objects these would
falsify MOND, but these claims remain
controversial primarily because of observational uncertainties:
ie inclination uncertainties, possibility of non-circular motions,
implied long virialization timescales.

\section{Near-isothermal pressure-supported systems}

These are self-gravitating objects supported not by rotation but
by the random motion
of their components having a velocity dispersion in a given system
that is roughly constant (isothermal).  These range from
giant molecular clouds within galaxies to globular star clusters
to dwarf spheroidal galaxies, luminous elliptical galaxies and
clusters of galaxies.  One might expect that they would be
described by isothermal solutions to the equation of hydrodynamics
which in spherical symmetry (and assuming isotropy of the velocity field) is
$${{\sigma_r^2}\over r}{d{ln\rho}\over {dlnr}} = -g \eqno(7)$$
where $\rho$ is the density, $\sigma_r$ is the (constant) radial component of the
velocity dispersion, and $g$ is the radial component of the gravitational
force.  In the Newtonian limit ($g = g_n$) the isothermal sphere has
infinite extent and mass with the density falling as $1/r^2$ and so
cannot represent a realistic description of these objects.  But with
MOND the asymptotic gravitational acceleration in the outer regions is
greater, falling as in eq.\ 3.  Thus
the equilibrium object is finite with a density (in low accelerations
regime) decreasing as $1/r^4$ and a finite mass given by
$${M\over{10^{11} M_\odot}} = {\Bigl({\sigma_r\over {100\, km/s}}\Bigr)}^4. \eqno(7)$$

In fact there is such observed relationship for bright elliptical
galaxies -- the Faber-Jackson relation relationship -- and this finds
a natural explanation in terms of MOND.  But MOND goes further as
a predictive theory:  in the context of MOND every near-isothermal
system should lie on roughly the same relationship; i.e., the
Faber-Jackson relationship should be universal (Sanders 2010).  An object with the
velocity dispersion of luminous galaxy (100-200 km/s) will have a mass on the order
of $10^{11}\, M_\odot$,
a typical galaxy mass;
if the  velocity dispersion is 1000 km/s, the mass would be $10^{14}\, M_\odot$,
the baryonic mass of a cluster of galaxies;  if the velocity dispersion is
5 to 10 km/s as for a globular star cluster, the mass is on the order of
$10^5/ M_\odot$, typical for such an object; and if the velocity dispersion is
2 - 5 km/s as observed in massive molecular clouds, the mass is
$10^3 - 10^4\, M_\odot$ as is observed for these objects.  The predication
of MOND is that the same $M\propto\sigma^4$ relationship apply to this
very wide range of astronomical objects ranging from sub-galactic clouds
to massive clusters of galaxies in so far as they are near-isothermal
systems supported by pressure.  Dark matter makes no such prediction.

But there is more.  The structure of a high surface density isothermal sphere in
MOND is basically that of a Newtonian sphere in the inner regions
with density falling as $1/r^2$.  But then at the radius where the
internal acceleration falls to $a_0$ the sphere is effectively truncated -- it does
not extend much beyond (density falling line $1/r^4$).  That means that
the internal acceleration of near isothermal spheres should be on the
order of $a_0$.  And this is true for such objects as we see in
Fig.\ 5.  Here we see that for this very wide range of near isothermal objects
it is true that the internal acceleration is on the order $a_0$.  Dark matter
offers no explanation of this observation, but it is a natural consequence
of MOND (Sanders \& McGaugh 2002).

\begin{figure}
\includegraphics[height=10cm]{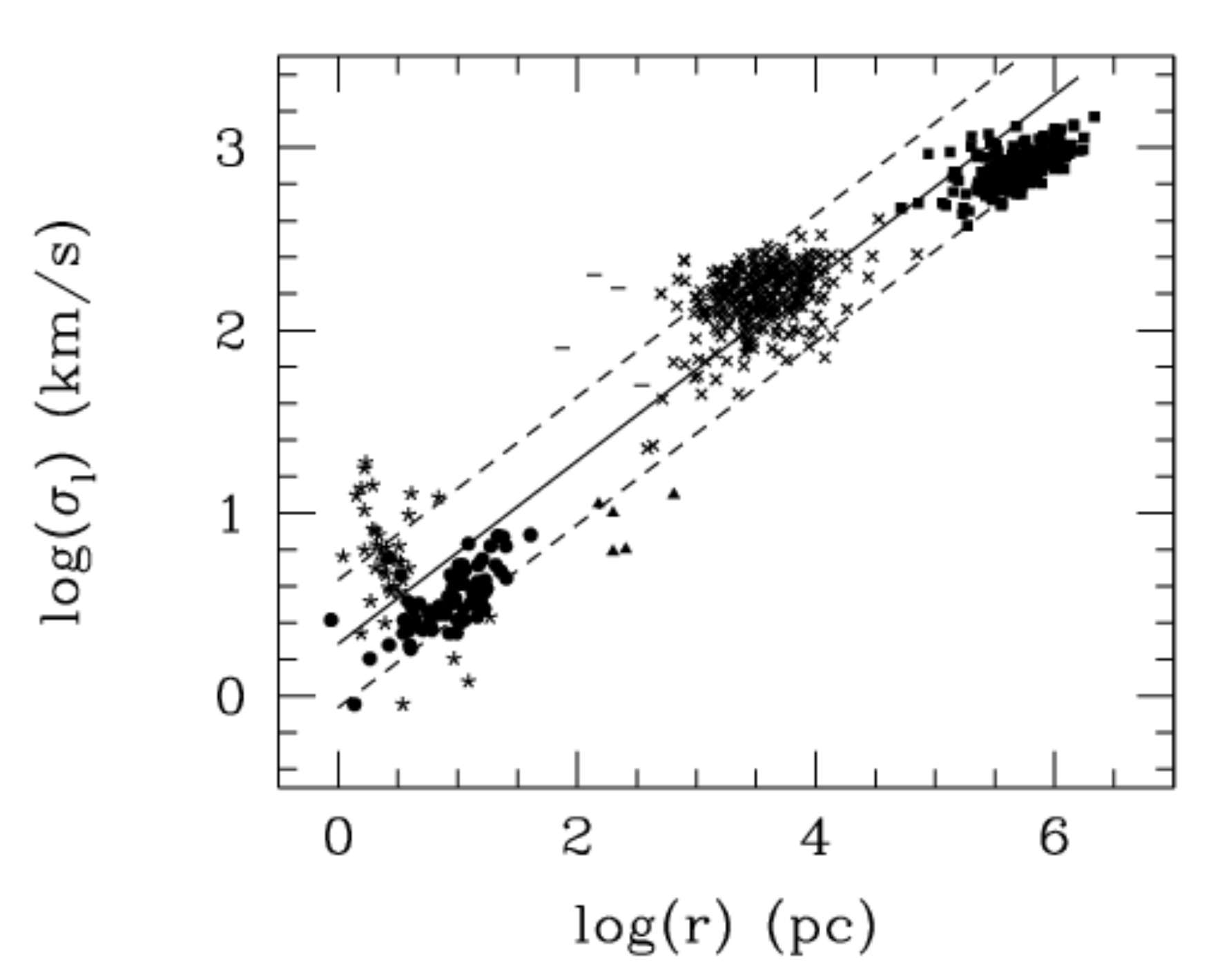}
\captionsetup{margin=10pt,font=small,labelfont=bf}
\caption{The velocity dispersion
size relation of self-gravitating objects
near
isothermal objects. The solid round points are molecular
clouds in the Milky Way, asterisks are globular clusters, triangles are
dwarf spheroidal galaxies, dashes are compact elliptial galaxies,
crosses are luminous elliptical galaxies, and the solid squares
are the giant clusters of galaxies.  The solid line corresponds to
$\sigma^2/r = a_0$ demonstrating that most of these objects have
an internal acceleration near $a_0$.  Points above the line would
have high surface brightness and little expected mass discrepancy.
Below the line are found low surface brightness objects with a large
expected discrepancy (Sanders \& McGaugh 2002).
}
\end{figure}

\section{Rotation curves of spiral galaxies}

Asymptotically flat rotation curves are a fundamental prediction of MOND.  But with
respect to reproducing rotation curves, MOND goes beyond this.  Milgrom first pointed
out that a general difference is expected between the rotation curves of
high surface brightness and low surface galaxies.  In LSB (low internal acceleration)
galaxies the discrepancy is present within the visible disk, and the
rotation curve will rise to the asymptotic value (consistent with the
Tully-Fisher relation, eq.\ 4).  But within HSB objects (high internal acceleration) the rotation
curve is effectively Newtonian and will fall in the outer galaxy to the
asymptotic value.  Thus, a simple rule:  rising rotation curves in LSB galaxies,
falling rotation curves in HSB galaxies.

With the advent (in the 1970s, see e.g. Bosma 1978)
of high quality observations of galaxies in the 21 cm line of HI
(extending well beyond the visible galaxy) this has been outstandingly confirmed.
In figure we see two examples in the neutral hydrogen rotation curves of
NGC 1560, an LSB galaxy,  NGC 2903 ,
an HSB galaxy (Begeman et al. 1991).  This in fact is a
general pattern of galaxy rotation curves.  But it is also obvious from this
analysis that even the details of these observed rotation curves are
well-accounted for by the MOND algorithm, as first emphasised by Begeman et al.
(1990) with one single universal new parameter, $a_0$, by the observed distribution
of baryons.

\begin{figure}
\begin{subfigure}[b]{0.9\textwidth}
\includegraphics[width=1\linewidth]{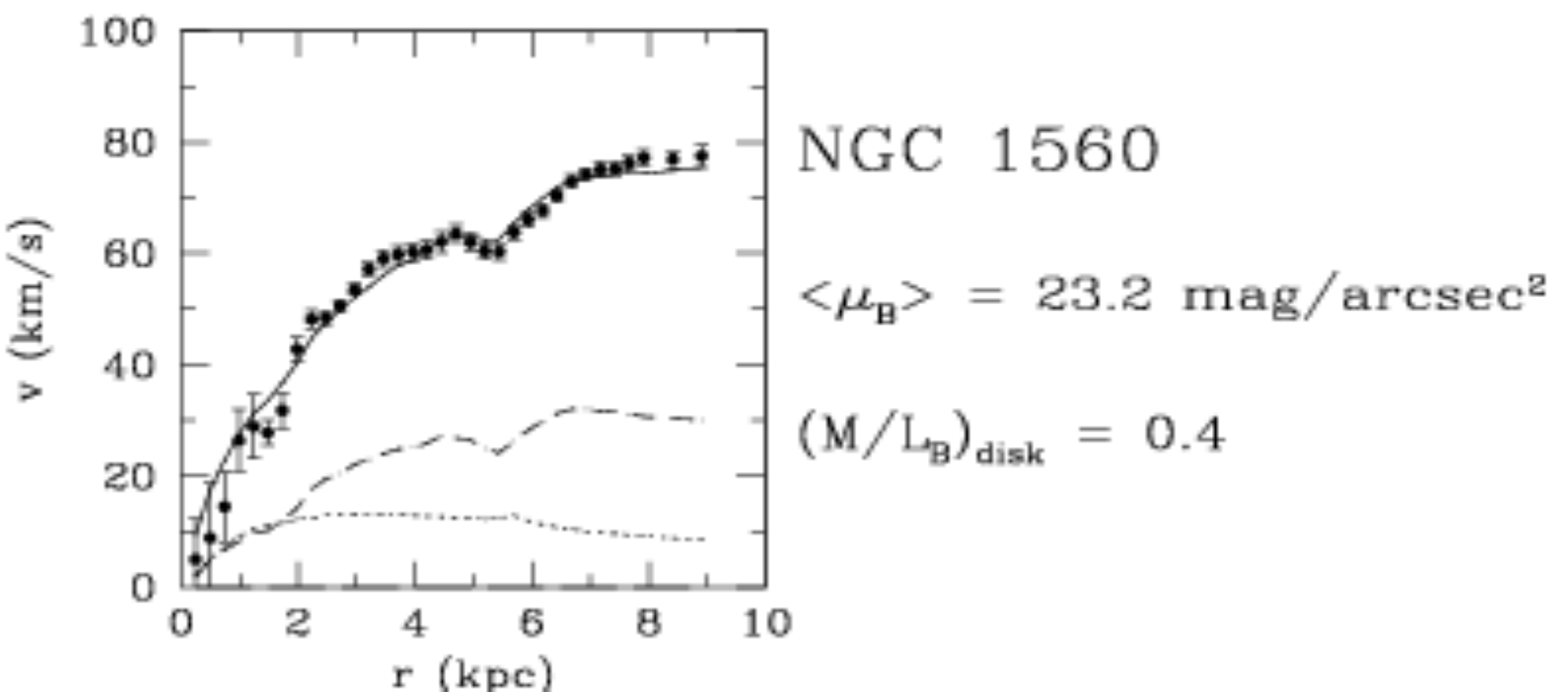}
\captionsetup{margin=10pt,font=small,labelfont=bf}
\caption{}
\end{subfigure}
\begin{subfigure}[b]{0.9\textwidth}
\includegraphics[width=1\linewidth]{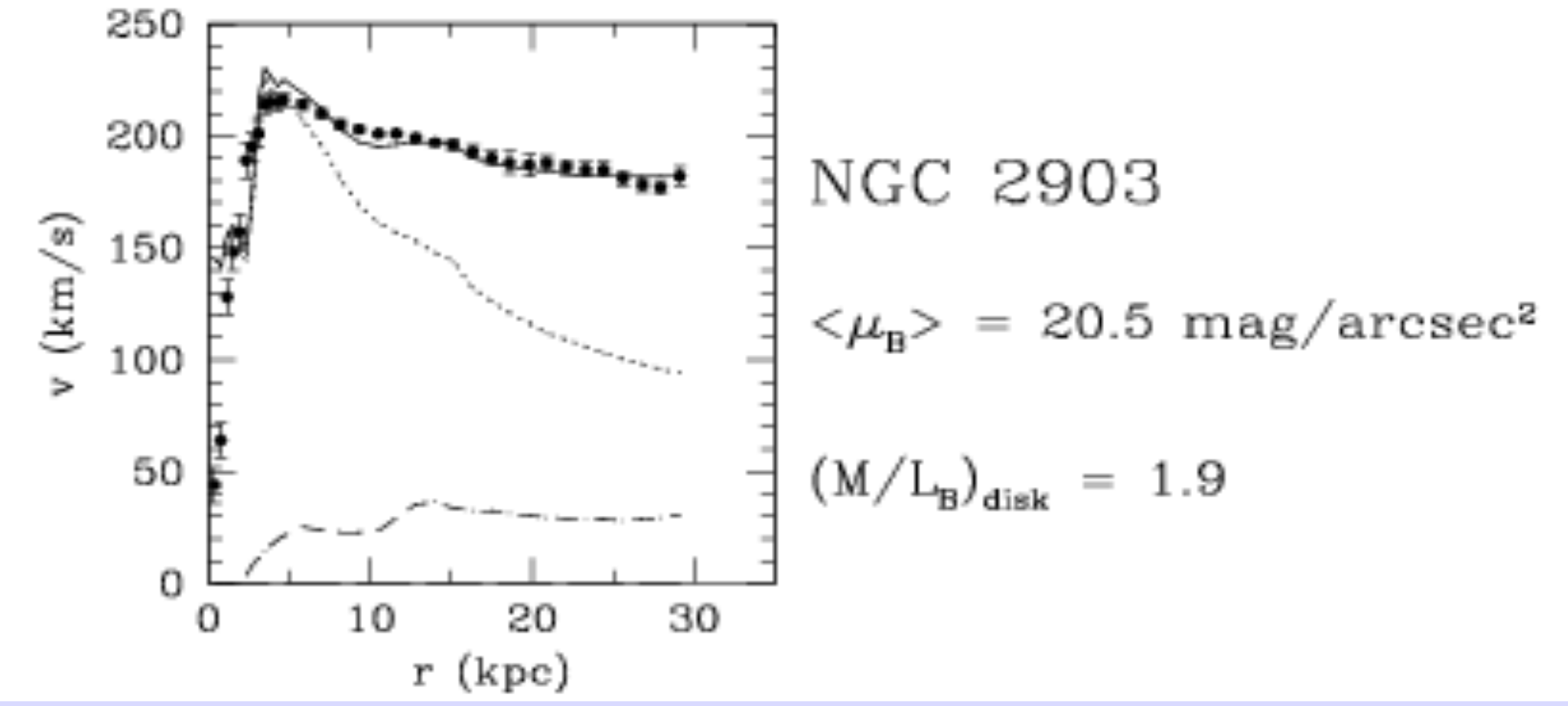}
\captionsetup{margin=10pt,font=small,labelfont=bf}
\caption{}
\end{subfigure}
\caption{Rotation curve $a$ of LSB galaxy ($\Sigma<\Sigma_c$)
and $b$ of HSB galaxy (($\Sigma>\Sigma_c$). Points with error bars
are the observations, dashed and dotted curves are the Newtonian
rotation curves of the baryonic components, gas and stars,
and the solid curve is the MOND curve with $a_0 = 1.2\times 10^{-8}$
$cm/s^2$ (Begeman, Broeils \& Sanders 1991)}
\end{figure}

However, in the earlier analyses there is an adjustable parameter which
may vary from galaxy to galaxy -- that is, the mass-to-light ratio of the
visible disk.  This quantity, assumed to be constant in a given galaxy,
determines the contribution of the stellar
disk to the gravity force within the object and is generally adjusted to achieve
the best agreement with the observations.
There are certain general constraints -- the fitted M/L should not
be much larger than is reasonable for the stellar population observed
in the Milky Way galaxy near the sun (on the order of unity in solar mass units),
otherwise
we are back to dark matter, and it should certainly not be negative.
In these initial studies these general constraints were met (there is no
a priori reason that they should be).  But in fact, MOND goes beyond these
generalities:  The implied mass-to-light ratios are completely
consistent with independent estimates of M/L on the basis of
models of stellar populations.

This is evident in Fig.\ 6 which is relevant to the rotation curves
for a sample of spiral galaxies
in the Ursa Major cluster and hence all at about the same distance
(Sanders \& Verneijen 1998).  The points here show the fitted mass-to-light
ratios of the galaxies (in terms of MOND) plotted against the B-V
colour index:  blue is to the left, red is to the right (Sanders \& McGaugh 2002).
The upper plot
is the B (blue) band M/L and the lower plot is the K band (near infrared)
M/L.  In the near-infrared the M/L values are
near constant but in the blue band, the redder galaxies have higher M/L
(recently formed stars have a lower M/L and emit a larger fraction of
their light in the blue).
The curves are not fits but are the theoretical M/L values for
a populations of stars having these average colours (Bell and de Jong 2001).
This is completely independent of the MOND estimated M/L values --
those required to achieve the optimal fits to the rotation curves using
the MOND algorithm.  The results are impressive considering that MOND has
no way of ``knowing" that redder galaxies should have a higher M/L.
If we had taken, a priori, the M/L from the population synthesis models
we could have achieved reasonable representations of the observed rotation
curves with no free parameters.

\begin{figure}
\includegraphics[height=10cm]{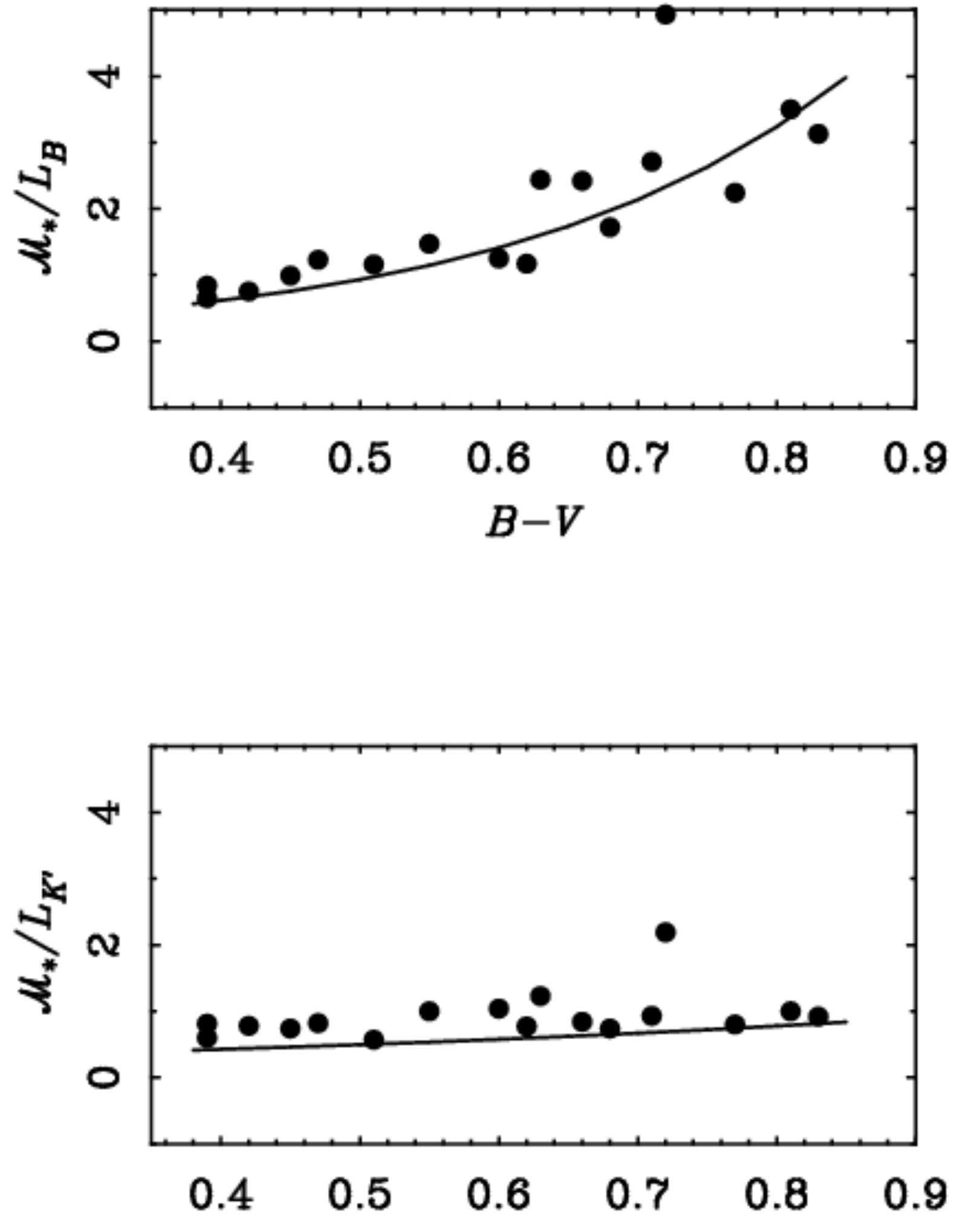}
\captionsetup{margin=10pt,font=small,labelfont=bf}
\caption{Points are values of M/L for spiral
galaxies following from MOND fits to rotation curves
in the UMa cluster (Sanders \& McGaugh 2002) plotted
against B-V colour index (blue to the left, red to the right)
in the blue band (top) and infrared (bottom).  The upper plot
is M/L in the blue band and in the lower plot, M/L in the
near infrared.  The solid curves are
M/L values predicted by population synthesis models (Bell and de Jong 2001)
}
\end{figure}

But true MOND predictions for rotation curves can be achieved for
a sample of galaxies where M/L is no longer a relevant parameter
at all; that is for gas-dominated dwarf galaxies.  In these galaxies,
the mass of neutral hydrogen is observed directly (in inferred from
the total 21 cm line emission).  It is found to completely dominate
the baryonic mass budget of the galaxies, i.e, the mass of gas
overwhelmingly exceeds the mass of of the visible stellar component.
Therefore, the mass-to-light ratio of the stellar component essentially
vanishes as an adjustable parameter.  There are complicating issues
with these small galaxies however.  The morphologies are
generally irregular which means that is difficult to estimate
the inclination of the plane in which the gas moves and thereby
correct the observed line-of-sight velocity to the true rotational
velocity.  In most cases there is clearly  evidence for warping
of the plane in which the gas moves and the additional complication of
non-motion due to deviations from perfect axial symmetry.  But even
given these uncertainties, the predicted MOND rotation curves
most often agree with the observed curves to high precision.

\begin{figure}
\begin{subfigure}[b]{0.9\textwidth}
\includegraphics[width=1\linewidth]{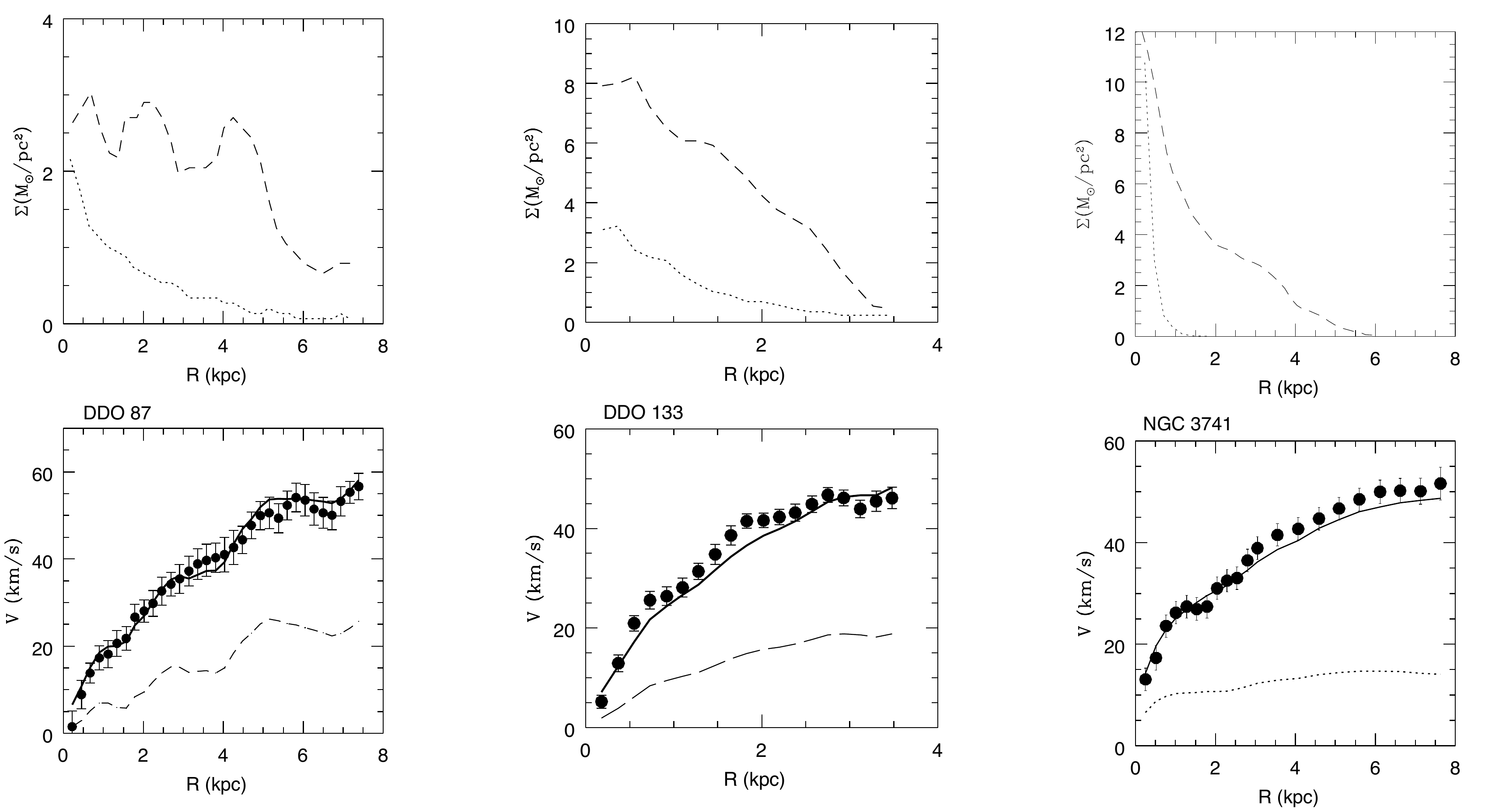}
\captionsetup{margin=10pt,font=small,labelfont=bf}
\caption{}
\end{subfigure}
\begin{subfigure}[b]{0.9\textwidth}
\includegraphics[width=1\linewidth]{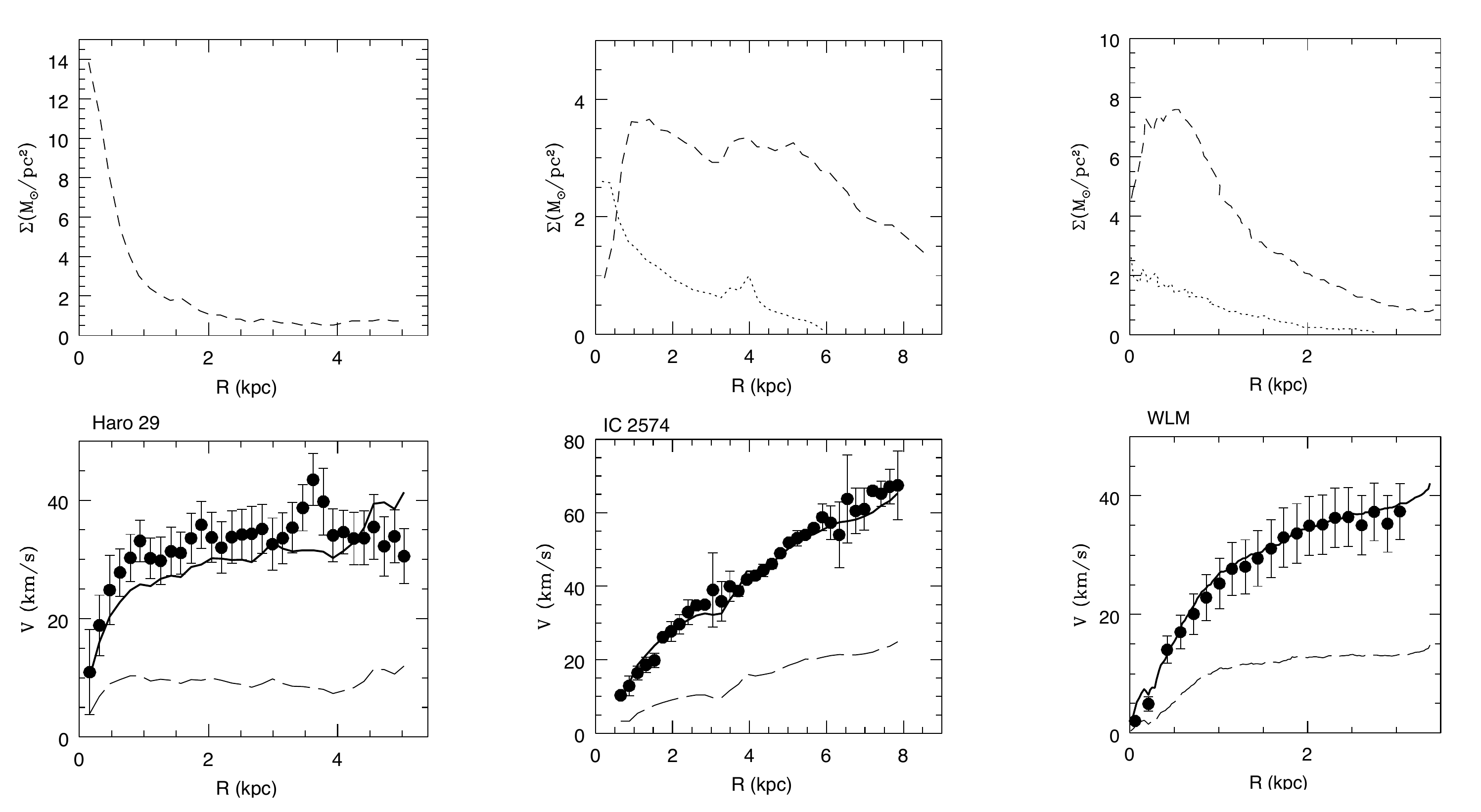}
\captionsetup{margin=10pt,font=small,labelfont=bf}
\caption{}
\end{subfigure}
\caption{Upper panels are baryonic surface density distribution (gas - dashed, dotted -stars))
and, lower panels, rotation curves of individual gas-dominated dwarf galaxies.  Points with
error bars are the observations; dashed curves are the Newtonian rotation curves
of baryonic components, and solid curves are the MOND rotation curves with
the standard value of $a_0$ (Sanders 2019).
}
\end{figure}
This is shown if Fig.\ 8 where we see six examples
of such gas-dominated objects from the observed sample
of Oh et al. 2015 considered by Sanders (2019).  For each
galaxy the top panel is the distribution of the baryonic
components (dashed - gas, dotted - stars) plotted as surface density
as a function of radius.  These have been taken directly
as given in the analysis of Oh et al. (2015) and not adjusted in any sense.
The bottom panel shows the rotation curve:  points are the
observations, dashed curve is the Newtonian rotation curve
of the baryonic components (mostly gas) and the solid curve
is the MOND curve calculated via eq. 3 with the standard value
of $a_0$.  The agreement of MOND with the observations is
noteworthy for most objects, particularly considering that
the MOND rotation curves are pure predictions from the
distribution of baryonic components.  Details of the observed
rotation curves in several cases are clearly
related to details in the observed gas distribution in the
presence of a large discrepancy.  It is difficult to imagine
that dark matter, as it is perceived to be, could achieve such
a level of successful prediction.

\section{DM and MD:  Competition or Completion}

In $\Lambda$CDM the basic gravitational or dynamical framework
is provided by general relativity which is a well-established
theory with no adjustable parameters and no established
contradictions on sub-galactic scales.  But to explain
cosmological observations general relativity alone is not
enough:  two unconventional sources must be added -- dark
energy and dark matter.  The nature or microphysics of these
two aethers is unknown and as long as this is so,
triumphalism over our present understanding of the Universe is premature.

It is well known that there are phenomenological problems
with the dark matter with respect to smaller scales where it
is assumed to cluster and make up the mass budget in systems
such as galaxies:  e.g., the ``core-cusp problem", the ``missing satellites".
These problems are generally brushed away because there is a
reductionist current in modern science which assigns priority to cosmology -- the
phenomenology of the entire Universe -- over the behaviour
of its mere constituents. The observational problems of galaxies are considered
minor and will be solved when there is a better understanding
of `` baryonic physics".

The essential problem of the dark matter
hypothesis concerns the regularities revealed by
galaxy scaling relations as well the details of
galaxy rotation curves.  With respect to the scaling
relations -- the Tully-Fisher, the radial acceleration
relation -- in the context of dark matter, these are
thought to emerge as aspects of galaxy formation; it seems
curious that such a precise relations can arise from what
is certainly a highly stochastic process with each individual
galaxy having its own history of dynamical evolution and merging.
With respect to rotation curves, the dark matter approach is to use the
observations to constrain the properties of halos.  The procedure
is to fit halo parameters (usually three) to the observations and to
apply semi-analytic repairs to fix details of the baryonic effects.
It may work, but the basic concept of dark halos cannot be falsified by
measurements of the force
distribution even in halo-dominated objects such as gas
rich dwarf galaxies where the total force is so obviously connected
to the gas distribution.  The exercise is essentially one of
post facto data fitting.

MOND subsumes scaling relations; they are an aspect of physical law
and not the random circumstances of formation -- hence their precision.
As an algorithm for calculating the rotation curves of spiral
galaxies from the observed distribution of baryonic matter, MOND is inherently
predictive and thus inherently falsifiable.  It is one simple
formula that involves one new universal parameter.
In so far as predictability in has value in science, this is a clear advantage,
and the fact that it works constitutes a severe challenge to the assumption that
dark matter dominates the mass budget of galaxies.  In addition, there is the
ubiquity of $a_0$ ($\approx cH_0$) in phenomena on sub-Hubble scales.  This parameter
appears as the acceleration below which the discrepancy is present in galaxies
(Fig.\ 1);
as a critical surface density within systems above which the discrepancy
is not apparent;
as the normalisation of the Tully-Fisher relation in spiral galaxies (Fig.\ 3); as the
normalisation
the Faber-Jackson relation in pressure supported systems from globular clusters
to clusters of galaxies;
as the typical internal acceleration in nearly-isothermal systems ranging
from sub-galactic self-gravitating molecular clouds to the great clusters of galaxy
(Fig.\ 5).
The ubiquitous presence of this critical acceleration in different settings
has no single explanation in the context of the CDM paradigm.

The strongest observational evidence for cosmic dark matter is
the matching of the
pattern of anisotropies in angular power spectrum in
cosmic microwave background emitted at the epoch
when that radiation decouples from matter at a redshift of about 1000.
That is not to say that this explanation is unique, but it is consistent
with a total mass abundance of dark matter five to six times greater than
that of baryons.  However, these observations do not extend down to the scale
of present self-gravitating systems where
it is most often assumed that the cosmic dark matter clusters and provides
the explanation for the mass discrepancy in galaxies.  Of course,
there is an efficiency in this explanation: only one sort of
mysterious substance need be invoked for cosmic and local phenomena
(less compelling now that
a second mysterious substance --  dark energy -- is also required).
But this idea does not work on local scales; it is falsified by
the algorithm which predicts rotation curves from the observed
distribution of baryons -- MOND.
Perhaps the
answer is more subtle.  Perhaps there is dualism involved --
that a substance (or a field) is implicated which behaves
like dark matter-dark energy
on cosmic scale but like modified dynamics on small scale.
This is a possibility for further consideration.

I am grateful to Moti Milgrom for useful comments on this manuscript.

\end{document}